# The mesoscopic dynamics of thermodynamic systems

D. Reguera, J.M. Rubí

*Departament de Física Fonamental, Facultat de Física, Universitat de Barcelona, Martí i Franquès, 1, 08028-Barcelona, Spain*

J.M.G. Vilar

*Computational Biology Center, Memorial Sloan-Kettering Cancer Center, 307 East 63rd Street, New York, NY 10021*

## Abstract

Concepts of everyday use like energy, heat, and temperature have acquired a precise meaning after the development of thermodynamics. Thermodynamics provides the basis for understanding how heat and work are related and with the general rules that the macroscopic properties of systems at equilibrium follow. Outside equilibrium and away from macroscopic regimes most of those rules cannot be applied directly. Here we present recent developments that extend the applicability of thermodynamic concepts deep into mesoscopic and irreversible regimes. We show how the probabilistic interpretation of thermodynamics together with probability conservation laws can be used to obtain Fokker-Planck equations for the relevant degrees of freedom. This approach provides a systematic method to obtain the stochastic dynamics of a system directly from its equilibrium properties. A wide variety of situations can be studied in this way, including many that were thought to be out of reach of thermodynamic theories, such as non-linear transport in the presence of potential barriers, activated processes, slow relaxation phenomena, and basic processes in biomolecules, like translocation and stretching.

## 1. Introduction

It is still a major challenge to understand how the wide variety of behaviors observed in every day experience, such as the usual processes of living systems, arise from the relatively simple and small set of laws that rule the microscopic world. There are a few exceptions. Systems in the condition of equilibrium strictly follow the rules of thermodynamics [1]. In such cases, the intricate behavior of large numbers of molecules can completely be characterized by a few variables that describe general average properties. It is possible to extend thermodynamics to situations that are at local equilibrium. This is the domain of validity of nonequilibrium thermodynamics [2]. Despite its generality, this theory has notorious limitations: it is applicable only to macroscopic systems, for which fluctuations are not important, and it operates within the linear response domain. Whereas the linear approximation is valid for many transport processes, such as heat conduction and mass diffusion, even in the presence of large gradients [3], [4] it is not appropriate for activated processes in which the system immediately enters the nonlinear domain. Small systems [5], such as single molecules in a thermal bath, in which fluctuations can be even larger than the mean values, are beyond the scope of that theory.

In this feature article, we present recent advances aimed at obtaining a simple and comprehensive description of the dynamics of nonequilibrium systems at the mesoscopic scale. These advances have provided not only a deeper understanding of the concept of local equilibrium but also a framework, reminiscent of nonequilibrium thermodynamics, to study fluctuations in nonlinear systems.

We show that the probabilistic interpretation of the density together with conservation laws in phase space and positiveness of global entropy changes sets the basis of a theory similar to nonequilibrium thermodynamics but of a much broader range of applicability. In particular, the fact that it is based on probabilities instead of densities allows us to consider mesoscopic systems and their fluctuations. The situations that can be studied with this formalism, named Mesoscopic Nonequilibrium Thermodynamics (MNET), include, among others, slow relaxation processes, barrier crossing dynamics, chemical reactions, entropic driving, and non-linear transport. These processes are in general non-linear. From the methodological point of view, given the equilibrium properties of a system, this theory provides a systematic and straightforward way to obtain its stochastic nonequilibrium dynamics in terms of Fokker-Planck equations.

In order to set grounds for the development of the formalism, we discuss first the basic nonequilibrium thermodynamic concepts used in a local equilibrium description of the systems. We then describe how MNET provides equilibrium statistical mechanics with a dynamics at the mesoscopic level. The key idea is to introduce statistical concepts in the framework of conservation laws of nonequilibrium thermodynamics. After introducing the general ideas, we discuss applications to nonlinear transport phenomena and activated processes that further illustrate the usefulness of the approach. Applications of the use of the concept of local equilibrium at the mesoscale are also discussed. We treat in particular inertial effects in diffusion, the slow relaxation dynamics and the translocation of a biomolecule. We conclude with a brief overview of other situations where this approach has been used and with a discussion of how far-away-from-equilibrium situations can be recast into a local equilibrium description.

## 2. Nonequilibrium thermodynamics

Nonequilibrium thermodynamics is a well established classical discipline built up on the grounds of two main hypotheses. First, the local equilibrium hypothesis [2] assumes that the thermodynamic variables defined in each subsystem of a conveniently partitioned system admit the same interpretation as in equilibrium. Second, the entropy production of any isolated system is always non-negative. The theory attributes the deviations from equilibrium to the presence of unbalanced forces, such as electric fields or gradients, which give rise to fluxes, such as electric or heat currents. Forces and fluxes are in a relation cause-effect compatible with the second law of thermodynamics and with the inherent symmetries, either macroscopic or microscopic. The dynamics follows from the local conservation laws for the thermodynamic field quantities, in which the fluxes are linear functions of the forces whose coefficients, the Onsager coefficients, satisfy reciprocity relations.

The scheme of nonequilibrium thermodynamics as sketched previously has been successfully used to analyze irreversible processes in systems of very different nature [6], [7]. To illustrate explicitly its method, we will apply it to the simple case of mass diffusion in one dimension. The first step is to compute the entropy production. At equilibrium, changes in entropy $S$ are given by the Gibbs equation

$$TdS = dE + pdV - \mu dM, \qquad (1)$$

in which the thermodynamic extensive variables are the internal energy $E$, the volume $V$, and the mass $M$ of the system, and the intensive variables the temperature $T$, the pressure $p$, and the chemical potential $\mu$. All these quantities may in general depend on time. For the sake of simplicity, we assume that the process takes place at constant temperature, energy, and volume. Local equilibrium here means that the Gibbs equation holds also for non-infinitely-slow changes in the variables. It is then possible to replace variations by time derivatives. Taking into account the spatial dependence through a density $\rho(x)$ in the spatial $x$–coordinate ($M = \int_V \rho(x)dx$), one obtains from Eq. (1) the entropy production,

$$T\frac{dS}{dt} = -\int \mu[x,\rho(x)]\frac{\partial \rho(x)}{\partial t}dx. \qquad (2)$$

The chemical potential depends explicitly on the density and, in principle, also on the spatial coordinate (as occurs for instance in the presence of an inhomogeneous external field). The conservation law

$$\frac{\partial \rho}{\partial t} = -\frac{\partial J}{\partial x}, \qquad (3)$$

after insertion into Eq. (2) and integration by parts with the assumption that the current vanishes at the boundaries, leads to

$$T\frac{dS}{dt} = -\int J\frac{\partial \mu}{\partial x}dx. \qquad (4)$$

The term $J$ denotes the flux of mass whose conjugated force is the gradient of the chemical potential. In the absence of nonlocal effects, the flux is proportional to the force

$$J = -L\frac{\partial \mu}{\partial x}, \qquad (5)$$

where $L \equiv L[x,\rho(x)]$ is the Onsager coefficient, which can in general depend on the thermodynamic variables as well as on the $x$-coordinate. For a chemical potential that does not depend explicitly on the spatial coordinate i.e., $\mu \equiv \mu[\rho(x)]$, Eq. (3) with Eq. (5) can be rewritten as the well-known diffusion equation [8]

$$\frac{\partial \rho}{\partial t} = \frac{\partial}{\partial x}D\frac{\partial \rho}{\partial x}, \qquad (6)$$

where the diffusion coefficient is $D \equiv L\frac{\partial \mu}{\partial \rho}$.

Nonequilibrium thermodynamics uses a set of local variables whose global counterparts coincide with the ones defined at equilibrium. This choice clearly restricts its application domain to the macroscopic level, at typical length scales much larger than any molecular size. In such a situation, the inherent molecular nature of matter can be ignored and one can adopt a continuum description in terms of a few conserved fields. Whereas this approximation has been extremely useful in the characterization of many irreversible processes, it is no longer valid for systems defined at the mesoscale when the typical time and length scales are such that the presence of fluctuations becomes relevant.

The linear character of the constitutive relations proposed by the theory should in principle be appropriate only when the magnitude of the gradients is small. In practice, linear relations have been proved to work well for many transport processes even in presence of large gradients. In contrast, for activated processes, the assumption of linearity fails. Whereas transport processes may perfectly operate in a linear regime, activated processes are genuinely nonlinear and cannot be analyzed with nonequilibrium thermodynamics. This fact has seriously limited the application of thermodynamic theories in areas like chemical kinetics and biophysics in which systems accede to nonequilibrium states via activation. A first attempt to overcome such difficulties was pioneered by Prigogine and Mazur [9], who extended thermodynamic concepts to irreversible phenomena in systems with internal degrees of freedom. Building up on these ideas, it is possible to develop a mesoscopic extension of the conventional nonequilibrium thermodynamics, as described in the following section, which is able to overcome such difficulties.

## 3. Thermodynamics and statistics

Reduction of the observational time and length scales of a system usually entails an increase in the number of degrees of freedom which have not yet equilibrated and that therefore exert an influence in the overall dynamics of the system. The nonequilibrated degrees of freedom will be denoted by $\gamma$ ($\equiv \{\gamma_i\}$) and may for example represent the velocity of a particle, the orientation of a spin, the size of a macromolecule or any coordinate or order parameter whose values define the state of the system in a phase space. The characterization at the mesoscopic level of the state of the system follows from $P(\gamma,t)$, the probability density of finding the system at the state $\gamma \in (\gamma, \gamma + d\gamma)$ at time $t$. The entropy of the system in terms of this probability can be expressed through the Gibbs entropy postulate [2,10]

$$S = S_{eq} - k_B \int P(\gamma,t) \ln \frac{P(\gamma,t)}{P_{eq}(\gamma)} d\gamma, \qquad (7)$$

where $S_{eq}$ is the entropy of the system when the degrees of freedom $\gamma$ are at equilibrium. If they are not, there is a contribution to the entropy that arises from deviations of the probability density $P(\gamma,t)$ from its equilibrium value $P_{eq}(\gamma)$, which is given by

$$P_{eq} \sim \exp\left(\frac{-\Delta W(\gamma)}{k_B T}\right) \qquad (8)$$

Here $\Delta W(\gamma)$ is the minimum reversible work required to create that state [11], $k_B$ is Boltzmann's constant, and $T$ is the temperature of the heat bath. Variations of the minimum work for a thermodynamic system are expressed as

$$\Delta W = \Delta E - T\Delta S + p\Delta V - \mu \Delta M + y\Delta Y + ..., \qquad (9)$$

where the term $y\Delta Y$ represents a generic work (electric, magnetic, surface work...) performed on the system, being $y$ the intensive parameter and $Y$ its conjugated extensive variable [12]. The expression of minimum reversible work (9) reduces to the different thermodynamic potentials by imposing the constraints that define those potentials [1]. For instance, for the case of constant temperature, volume and number of particles, the minimum work corresponds to the Helmholtz free energy $A$. The statistical mechanics definition of the entropy $S$ is therefore the key to connect thermodynamics with both the mesoscopic description in terms of the probability distribution $P(\gamma,t)$ and the equilibrium behavior of the system. The combination of the statistical definition of the entropy with the systematic methodology of nonequilibrium thermodynamics results in a powerful framework to describe the kinetics of a wide class of systems. This framework is outlined in the next section.

## 4. Thermodynamics and stochastic dynamics

To describe the dynamics of the mesoscopic degrees of freedom, the starting point is the statistical mechanics definition of the entropy given through the Gibbs entropy postulate [2]. Taking variations in Eq. (7), one obtains

$$\delta S = -k_B \int \delta P(\gamma,t) \ln \frac{P(\gamma,t)}{P_{eq}(\gamma)} d\gamma. \tag{10}$$

The evolution of the probability density in the $\gamma-$space is governed by the continuity equation

$$\frac{\partial P}{\partial t} = -\frac{\partial J}{\partial \gamma}, \tag{11}$$

where $J(\gamma,t)$ is a current or density flux in the internal space which has to be specified. Its form can be obtained by taking the time derivative in equation (10) and by using the continuity equation (11) to eliminate the probability time derivative. After a partial integration, one then arrives at

$$\frac{dS}{dt} = -\int \frac{\partial}{\partial \gamma} J_S d\gamma + \sigma,$$

where

$$J_S = k_B J \ln \frac{P}{P_{eq}}$$

is the entropy flux, and

$$\sigma = -k_B \int J(\gamma,t) \frac{\partial}{\partial \gamma} \left( \ln \frac{P(\gamma,t)}{P_{eq}(\gamma)} \right) d\gamma, \tag{12}$$

is the entropy production.
In this scheme, the thermodynamic forces are identified as the gradients in the space of mesoscopic variables of the logarithm of the ratio of the probability density to its equilibrium value. We will now assume a linear dependence between fluxes and forces and establish a linear relationship between them

$$J(\gamma,t) = -k_B L(\gamma, P(\gamma)) \frac{\partial}{\partial \gamma} \left( \ln \frac{P(\gamma,t)}{P_{eq}(\gamma)} \right), \tag{13}$$

where $L(\gamma, P(\gamma))$ is an Onsager coefficient, which in general depends on the state variable $P(\gamma)$ and on the mesoscopic coordinates $\gamma$. To derive this expression, locality in $\gamma-$space has also been taken into account, for which only fluxes and forces with the same value of $\gamma$ become coupled. A similar expression for the current was derived from a free energy functional for the case of polymer dynamics [13].
The resulting kinetic equation follows by substituting Eq. (13) back into the continuity equation (11):

$$\frac{\partial P}{\partial t} = \frac{\partial}{\partial \gamma}\left(DP_{eq}\frac{\partial}{\partial \gamma}\frac{P}{P_{eq}}\right), \tag{14}$$

where the diffusion coefficient is defined as

$$D(\gamma) \equiv \frac{k_B L(\gamma, P)}{P}. \tag{15}$$

This equation, which in view of Eq. (8) can also be written as

$$\frac{\partial P}{\partial t} = \frac{\partial}{\partial \gamma}\left(D\frac{\partial P}{\partial \gamma} + \frac{D}{k_B T}\frac{\partial \Delta W}{\partial \gamma}P\right), \tag{16}$$

is the Fokker-Planck equation for the evolution of the probability density in $\gamma$-space. Under the conditions for which the minimum work is given by the Gibbs free energy $G$, $\Delta W \equiv \Delta G = \Delta H - T\Delta S$, where $H$ is the enthalpy, this equation transforms into the Fokker-Planck equation for a system in the presence of a free energy barrier:

$$\frac{\partial P}{\partial t} = \frac{\partial}{\partial \gamma}\cdot\left(D\frac{\partial P}{\partial \gamma} + \frac{D}{k_B T}\frac{\partial \Delta G}{\partial \gamma}P\right). \tag{17}$$

Other cases of interest concern different thermodynamic potentials. For instance, a particularly interesting situation is the case of a purely entropic barrier, often encountered in soft-condensed matter and biophysics, which will be discussed in detail in Sect. 5.1.

It is important to stress that MNET provides a simple and direct method to determine the dynamics of a system from its equilibrium properties. In particular, by knowing the equilibrium thermodynamic potential of a system in terms of its relevant variables, one could easily derive the general form of the kinetics. The method proposed thus offers a general formalism able to analyze the dynamics of systems away from equilibrium. In the following section we will illustrate its applicability by means of some examples.

The scheme presented can be put in closer connection with nonequilibrium thermodynamics concepts. To that end, the crucial idea is the generalization of the definition of the chemical potential to account for this additional mesoscopic variables. We may then assume that the evolution of these degrees of freedom is described by a diffusion process and formulate the corresponding Gibbs equation

$$\delta S = -\frac{1}{T}\int \mu(\gamma)\delta P(\gamma, t)d\gamma, \tag{18}$$

which resembles the corresponding law proposed in nonequilibrium thermodynamics for a diffusion process in terms of the mass density of particles. Here $\mu(\gamma)$ plays the role of a generalized chemical potential conjugated to the distribution function $P(\gamma, t)$. Comparison of the Gibbs equation (18) with Eq. (10), where the variations of the equilibrium entropy are given by

$$\delta S_{eq} = -\frac{1}{T}\int \mu_{eq}\delta P(\gamma, t)d\gamma, \tag{19}$$

and $\mu_{eq}$ is the value of the chemical potential at equilibrium, yields the identification of the generalized chemical potential as

$$\mu(\gamma, t) = k_B T \ln \frac{P(\gamma, t)}{P_{eq}(\gamma)} + \mu_{eq}, \tag{20}$$

or alternatively, using Eq. (8),

$$\mu(\gamma, t) = k_B T \ln P(\gamma, t) + \Delta W. \tag{21}$$

In this reformulation, the "thermodynamic force" driving this general diffusion process is $T^{-1}\partial\mu/\partial\gamma$, and the entropy production is given by

$$\sigma = -\frac{1}{T}\int J\frac{\partial\mu}{\partial\gamma}d\gamma \tag{22}$$

By comparison of the previous equation with eq. (4), it is clear that the evolution in time of the system mimics a generalized diffusion process over a potential landscape in the space of mesoscopic variables. This landscape is conformed by the values of the equilibrium energy associated to each configuration $\gamma$. The treatment of a diffusion process in the framework of nonequilibrium thermodynamics can then be extended to the case in which the relevant quantity is a probability density instead of a mass density. The relation between entropy and stochastic dynamics has also been discussed in a different context in refs14 and15.

## 5. Applications

Transport at the mesoscale is usually affected by the presence of forces of different nature: direct interactions between particles, hydrodynamic interactions mediated by the solvent and excluded volume effects. The presence of such diversity of forces has a direct implication in the form of the energy landscape, which may exhibit a great multiplicity of local minima separated by barriers. Transport at those scales presents two main characteristics: it is intrinsically nonlinear and it is influenced by the presence of fluctuations, external driving forces, and gradients.

The mesoscopic nonequilibrium thermodynamics theory proposed can be used to infer the general kinetic equations of a system in the presence of potential barriers, which in turn can be used to obtain the expressions for the current of particles and the diffusion coefficient. Both quantities are related to the two first moments of the distribution function and are accessible to the experiments. In this section we illustrate the application of the theory to different representative situations. We will discuss nonlinear transport processes in which the dynamics is influenced by the presence of entropic forces and activated processes in both homogeneous and inhomogeneous environments.

*1. Kinetic processes in the presence of entropic forces*

Entropy and entropic forces play major roles in soft condensed matter and biophysics where typical energies are of the order of the thermal energy. The theory introduced in the previous section can be easily applied to account for this situation. In the case of entropic forces, the minimum reversible work introduced in eq. (9) is $\Delta W = -T\Delta S$. The corresponding kinetic equation is

$$\frac{\partial P}{\partial t} = \frac{\partial}{\partial\gamma}\left(D\frac{\partial P}{\partial\gamma} - \frac{D}{k_B}\frac{\partial\Delta S}{\partial\gamma}P\right), \tag{23}$$

where now the $\gamma$-coordinate includes the coordinates necessary to characterize the evolution of the system under the influence of the entropic potential. This equation constitutes the starting point in the study of transport processes in the type of confined systems that are very often encountered at sub-cellular level and in microfluidic applications. The basic situation of the motion of a Brownian particle in an enclosure of varying cross-section was analyzed using this perspective in Ref. [16]. The main idea is that the very complicated boundary conditions of the diffusion equation in irregular channels can be greatly simplified by the introduction of an entropic potential that accounts for the space accessible for the diffusion of the Brownian particle. In the case of a 3D pore of cross-section $A(x)$, the entropic potential can be easily calculated by contracting the 3D description and by retaining only the coordinate $x$. The resulting 1D equilibrium distribution $P_{eq}(x)$ is

$$P_{eq}(x) = \int P_0 dy dz = P_0 A(x) \qquad (24)$$

where $P_0$ is the probability distribution in the absence of potential, assumed constant. From this result and Eq. (8) we then infer the expression for the entropic potential $\Delta S(x) = -k_B \ln A(x)$. By substitution of this potential in eq. (23) one obtains

$$\frac{\partial P}{\partial t} = \frac{\partial}{\partial x}\left( D(x)\frac{\partial P}{\partial x} - \frac{D(x)}{A(x)}\frac{\partial A(x)}{\partial x} P \right), \qquad (25)$$

which is known as Fick-Jacob equation [17]. As a result of the contraction, the effective diffusion coefficient depends on the coordinate. Using scaling arguments one finds [16]

$$D(x) = D_0 \frac{1}{(1+y'(x)^2)^\alpha} \qquad (26)$$

where $D_0$ is the molecular diffusion coefficient, $y(x)$ defines the shape of the enclosure and $\alpha$ is a scaling exponent whose value is 1/3 for the 2D case and 1/2 for the 3D case. The current of particles through the pore and the effective diffusion coefficient depend strongly on the entropic forces related to the shape of the container in which particles move and can be computed from the Fick-Jacob equation. The results obtained agree with the exact solution of the 3D diffusion equation over a wide range of conditions (see Fig. 1).

Through the basic scenario presented, we can analyze the effects of entropic forces in the dynamics of the system. Entropic forces are present in many situations, such as the motion of macromolecules through pores, phoretic effects, transport through ion channels, protein folding, and in general in the dynamics of small confined systems [18,19].

## 2. Activated processes

Activated processes are those that need a finite energy to proceed and change the system from one state to another. The paradigm of activated processes is the crossing of a free energy barrier that separates two well-differentiated states that lie at the local minima at each side of the barrier. The system needs to acquire energy to surmount the barrier. Once the barrier is crossed, energy is released. Processes like thermal emission in semiconductors, chemical reactions, adsorption, nucleation, and active transport through biological membranes, share these features and, therefore, are generically referred to as activated processes [20].

It is important to emphasize the essential difference between activated processes and the linear processes described by nonequilibrium thermodynamics. The latter constitute the response to the application of an external force or gradient and may emerge even at very low values of the applied force, in the linear response regime.

Contrarily, the regime in which activated processes may take place is basically nonlinear. In this context, we can contrast the linear Fourier, Fick, or Ohm laws, in which the corresponding currents are proportional to the conjugated thermodynamic forces or gradients, with the exponential laws appearing in activated processes.

Let us consider a general process for which a system passes from state 1 to state 2 via activation. Instances of this process can be a chemical reaction in which a substance transforms into another, an adsorption process in which the adsorbing particle goes from the physisorbed to the chemisorbed state, or a nucleation process in which the metastable liquid transforms into a crystal phase. Nonequilibrium thermodynamics describes the process only in terms of the initial and final positions and is valid only in the linear response regime [2]. If we consider the process at shorter time scales, the state of the system progressively transforms by passing through successive molecular configurations. These different configurations can be characterized by a "reaction coordinate" $\gamma$. In this situation, one may assume that this reaction coordinate undergoes a diffusion process through a potential barrier separating the initial from the final states. The local entropy production is (see also Eq. (22))

$$\sigma(\gamma,t) = -\frac{1}{T} J \frac{\partial \mu}{\partial \gamma} \qquad (27)$$

from which we can infer the linear law

$$J(\gamma,t) = -\frac{L}{T} \frac{\partial \mu}{\partial \gamma} \qquad (28)$$

where the chemical potential is, as in Eq. (21), given by

$$\mu = k_B T \ln P + \Phi \qquad (29)$$

with $\Phi(\gamma)$ being the potential in terms of the reaction coordinate (see Fig. 2). Following the previous approach, we can obtain the Fokker-Planck equation for the dynamics of $\gamma$

$$\frac{\partial P}{\partial t} = \frac{\partial}{\partial \gamma} \left( b(\gamma) P(\gamma) \frac{\partial \mu(\gamma)}{\partial \gamma} \right), \qquad (30)$$

where $b(\gamma)$ is a mobility in the $\gamma$-space. This equation describes the dynamics of $\gamma$ for an arbitrary potential and at any value of the temperature.

It is often the case that, at the time scales of interest, the system is mostly found in the states 1 and 2, which correspond to the minima at $\gamma_1$ and $\gamma_2$, respectively. The probability distribution is strongly peaked at these values and almost zero everywhere else. This happens when the energy barrier is much higher than the thermal energy and intra-well relaxation has already taken place. Using MNET, we will show that the Fokker-Planck description, under these conditions, leads to a kinetic equation in which the net reaction rate satisfies the mass action law.

The current given in (28) can be rewritten in terms of the local fugacity defined along the reaction coordinate $z(\gamma) \equiv \exp \mu(\gamma)/k_B T$ as

$$J = -k_B L \frac{1}{z} \frac{\partial z}{\partial \gamma}, \qquad (31)$$

which can be expressed as

$$J = -D \frac{\partial z}{\partial \gamma}, \qquad (32)$$

where $D = k_B L/z$ represents the diffusion coefficient. We now assume $D$ constant and integrate from 1 to 2, obtaining

$$\bar{J} \equiv \int_1^2 J d\gamma = -D(z_2 - z_1) = -D(\exp\frac{\mu_2}{k_B T} - \exp\frac{\mu_1}{k_B T}). \quad (33)$$

This equation can alternatively be expressed as

$$\bar{J} = J_0 \left(1 - e^{A/k_B T}\right),$$

where $\bar{J}$ is the integrated rate, $J_0 = D\exp(\mu_1/k_B T)$ and $A = \mu_1 - \mu_2$ is the corresponding affinity. We have then shown that MNET leads to nonlinear kinetic laws. Remarkably, it is possible to move from a linear continuous to a nonlinear discrete system; that is to say, a Fokker-Planck equation, linear in probabilities and in the gradient of $\mu[\gamma, P(\gamma)]$, accounts for a non-linear dependence in the affinity. This scheme has been successfully applied to different classical activated processes, like chemical reactions [21], adsorption [22], thermal emission in semiconductors [23], or nucleation [24], to obtain the corresponding kinetic laws.

## 3. Activated processes in inhomogeneous systems

In many practical instances, the activated process takes place in the presence of gradients of thermodynamic or hydrodynamic quantities. In this situation the bath has its own nonequilibrium dynamics which is coupled to that of the system. This is what happens for example in inhomogeneous nucleation where the germs emerge and growth in a nonequilibrium metastable liquid. The inhomogeneities in the bath may exert a significant influence in the expressions of the activation rates. The mesoscopic theory we have proposed describes the coupled evolution of the system and the bath by providing the hydrodynamic equations for the bath together with the kinetic equation. To illustrate the application of the method we will discuss here nucleation when the metastable phase is subjected to a temperature gradient $\nabla T$ and to a velocity gradient $\nabla \mathbf{v}$.

The case of a temperature gradient was discussed in detail in Ref. [25]. The kinetics of formation of clusters in the presence of density inhomogeneities and a temperature gradient is described by the kinetic equation

$$\frac{\partial f_c}{\partial t} = \nabla \cdot \left(D_0 \nabla f_c + D_{th} \frac{\nabla T}{T} f_c\right) + \frac{\partial}{\partial n} J_n. \quad (34)$$

which can be obtained by following the procedure indicated in Sect. 4. In this equation $f_c(n, x, t)$ is the number fraction of clusters containing $n$ molecules at a point $x$ of the sample at time $t$, $D_0$ and $D_{th}$ are the spatial and thermal diffusion coefficients, respectively, and $J_n$ is the rate of formation of clusters of size $n$. Therefore, the first term on the right hand side of Eq. (33) accounts for the effects of diffusion and temperature gradients in the process of nucleation.

For the case of a velocity gradient, the corresponding kinetic equation was obtained in [26]

$$\frac{\partial f_c}{\partial t} = -\nabla \cdot (f_c \mathbf{v}_0) + \nabla \cdot \left(\vec{\vec{D}} \cdot \nabla f_c\right) + \frac{\partial}{\partial n} J_n, \quad (35)$$

where $\mathbf{v}_0$ is the velocity field, and

$$\vec{\vec{D}} = D_0 \left[ \vec{\vec{1}} - \left( \frac{\vec{\vec{\eta}}_B}{p} \cdot \nabla \mathbf{v}_0 \right)^0 \right], \tag{36}$$

is an effective diffusion coefficient of tensorial nature due to the breaking of the isotropy of the system by the shear flow. The upper $0$ means symmetric part of the tensor. The presence of the shear rate introduces a correction to the Stokes-Einstein diffusion coefficient $D_0 \vec{\vec{1}}$, with $\vec{\vec{1}}$ the unit matrix, which depends on the strength of the shear rate, the Brownian viscosity $\vec{\vec{\eta}}_B$ related to the stresses exerted by the Brownian motion of the clusters, and the pressure $p$. Eq. (35) thus accounts for the effects of the shear flow on the kinetics of nucleation. These effects are accordingly more pronounced at high shear rates and close to the glass transition when the viscosity increases significantly. This situation is frequently encountered in polymer crystallization which normally proceeds at very large values of the shear rate. Experiments show spherical forms of the clusters when they grow at rest and elongated forms when the metastable phase is sheared (see Fig. 3), in agreement with the anisotropy of the diffusion process [27].

The previously described situations can in general be found in different systems undergoing activation dynamics under inhomogeneous conditions, as in ion channels, protein binding kinetics and diverse macromolecular transport processes, and illustrate the important influence that the bath exerts in the evolution of the system.

## 6. Local equilibrium at the mesoscale

Bringing together thermodynamics and stochastic dynamics to describe mesoscopic systems relies on the assumption of local equilibrium. Nonequilibrium thermodynamics considers that there is local equilibrium when a system can be subdivided into smaller subsystems that look homogenous and yet macroscopic. In addition, the thermodynamic variables that characterize each of these subsystems have to evolve sufficiently slow compared to the microscopic time scales.

The local equilibrium condition can be interpreted in a more general context, which accounts for time scales in which not all the fast variables have relaxed. In this new interpretation, systems which are not in local equilibrium could equilibrate locally when the non-equilibrated fast variables are incorporated into the thermodynamic description. This description requires the increase in the number of variables and the resulting increase of the configurational space. Thus, systems outside equilibrium can be brought to local equilibrium in terms of the extended set of variables. This possibility will be illustrated in the following subsections through examples of very different nature: mass diffusion in the presence of inertial effects, relaxation phenomena in glassy systems, and the nonequilibrium translocation of a biomolecule through a pore.

*1. Local equilibrium and inertial effects in diffusion*

Inertial effects should be taken into account in diffusion when changes in the spatial density occur at a time scale comparable with the time the velocities of the constituent elements need to relax to equilibrium. In such a situation, the local equilibrium assumption does not hold; at each point the velocity field is still relaxing towards its equilibrium state and the entropy production depends on the particular form of the velocity distribution. In this case, conventional nonequilibrium thermodynamics is not valid.

We will show that local equilibrium can be restored if the space of variables is enlarged by incorporating the velocity as an additional coordinate; both the spatial coordinate, $x$, and velocity coordinate, $v$, are needed to completely specify the state of the system. In this case, we can proceed as indicated before, by considering the subsystem at local equilibrium in the two-dimensional space $\gamma = (x, v)$.

From equilibrium statistical mechanics we obtain that the chemical potential is given by

$$\mu(x,v) = \Phi(x) + \frac{1}{2}v^2 + kT \ln P(x,v), \qquad (37)$$

where the second term is the kinetic energy per unit mass of the constituent elements and $\Phi(x)$ their potential energy.

From the probability conservation law we obtain that the Fokker-Planck equation follows

$$\frac{\partial P}{\partial t} = -\frac{\partial J_x}{\partial x} - \frac{\partial J_v}{\partial v}, \qquad (38)$$

with

$$J_x = -L_{xx}\frac{\partial \mu}{\partial x} - L_{xv}\frac{\partial \mu}{\partial v}, \qquad (39)$$

$$J_v = -L_{vx}\frac{\partial \mu}{\partial x} - L_{vv}\frac{\partial \mu}{\partial v}, \qquad (40)$$

where $L_{xx}, L_{xv}, L_{vx}$, and $L_{vv}$ are the Onsager coefficients.

We will now study how these general Onsager coefficients are constrained by the physics of diffusion processes. First, the flux of probability in $x$-coordinate space, $\tilde{J}_x(x) \equiv \int_{-\infty}^{\infty} v P(x,v) dv$, has to be recovered from the flux in the ($x,v$)-space by contracting the velocity coordinate: $\tilde{J}_x(x) = \int_{-\infty}^{\infty} J_x(x,v) dv$. Therefore,

$$\int_{-\infty}^{\infty} v P \, dv = -\int_{-\infty}^{\infty} \left( L_{xx}(\frac{\partial \Phi}{\partial x} + \frac{kT}{P}\frac{\partial P}{\partial x}) + L_{xv}(v + \frac{kT}{P}\frac{\partial P}{\partial v}) \right) dv. \qquad (41)$$

where we have used Eqs. (37) and (39). Since $P(x,v)$ can take any arbitrary form, the last equality holds if and only if $L_{xx} = 0$ and $L_{xv} = -P$. Second, the positiveness of the entropy production, $\sigma = (L_{xv} + L_{vx})\frac{\partial \mu}{\partial x}\frac{\partial \mu}{\partial v} + L_{vv}(\frac{\partial \mu}{\partial v})^2$, implies that $L_{xv} = -L_{vx}$, which coincides with the Onsager relations [28]. Thus, the only undetermined coefficient is $L_{vv}$, which can depend explicitly on $x$ and $v$.

Previous equations can be rewritten in a more familiar form by identifying the Onsager coefficients with macroscopic quantities. In this way, with $L_{vv} = P/\tau$, the fluxes read

$$J_x = \left( v + \frac{D}{\tau}\frac{\partial}{\partial v} \right) P, \qquad (42)$$

$$J_v = -\left( \frac{1}{\tau}\frac{\partial \Phi}{\partial x} + \frac{D}{\tau}\frac{\partial}{\partial x} + \frac{v}{\tau} + \frac{D}{\tau^2}\frac{\partial}{\partial v} \right) P, \qquad (43)$$

where $D \equiv kT\tau$ and $\tau$ are the diffusion coefficient and the velocity relaxation time, respectively. The equation for the density is given by [29,30]

$$\frac{\partial P}{\partial t} = -\frac{\partial}{\partial x}vP + \frac{\partial}{\partial v}\left(\frac{1}{\tau}\frac{\partial \Phi}{\partial x} + \frac{v}{\tau} + \frac{D}{\tau^2}\frac{\partial}{\partial v}\right)P \tag{44}$$

which describes the influence of inertial effects in the diffusion.

The previous example illustrates that when the variables that would increase the entropy during the processes of interest are considered explicitly, the local equilibrium assumption is valid in the space defined by such variables. By incorporating all the relevant variables $\gamma$ into the description, it is possible to use MNET to study systems that are not as close to equilibrium as required by standard nonequilibrium thermodynamics.

Let us study in more detail the concept of local equilibrium in an extended space. For simplicity, we will consider the case $\Phi(x) = 0$. The condition of equilibrium is characterized by the absence of dissipative fluxes; that is to say, by $J_x = 0$ and $J_v = 0$. Therefore, from Eq. 42 we obtain the well-known equilibrium result that the velocity distribution is Gaussian with variance proportional to the temperature. If deviations from equilibrium are small ( $J_x \neq 0$ and $J_v = 0$ ), the local equilibrium hypothesis holds. This is the domain of validity of Fick's law,

$$J_x = -D\frac{\partial P}{\partial x}, \tag{45}$$

which is obtained directly from the equations for the fluxes. In this case, the distribution of velocities is still Gaussian, as in equilibrium, but now centered at $\bar{v}(x) = \int_{-\infty}^{\infty} vP(x,v)dv = \int_{-\infty}^{\infty} \frac{D}{\tau}\frac{\partial P(x,v)}{\partial x}dv$ and the variance of the distribution is related to the temperature. When local equilibrium holds in the $\gamma-$ space but not in the $x-$ space ( $J_x \neq 0$ and $J_v \neq 0$ ), the velocity distribution is not longer constrained to have a Gaussian form. In Fig 4 we have represented the probability distribution function solution of Eq. (43) for the case in which an imposed concentration gradient keeps the system outside equilibrium. When the velocity distribution relaxes very fast to the Maxwellian equilibrium distribution, the probability distribution is a local Gaussian. On the contrary, for larger values of the velocity relaxation time the Gaussian nature of the distribution function is lost. These non-Gaussian forms have been recently found in experiments performed with single-molecules [31] and glasses [32, 33] under nonequilibrium situations.

## 2. Translocation of a biomolecule

Many biological processes involve the translocation of proteins or nucleic acids through pores or channels. One of the most common examples is the translocation of proteins from the cytosol to the endoplasmatic reticulum or the entry of the DNA of a bacteriophage into the cell. The simplest mechanism of translocation of a biomolecule is by simple diffusion. However, this mechanism is very slow, and in many cases there are some proteins that facilitate the entry of the biomolecule by binding reversibly to it. The role commonly assumed to be played by these proteins was to rectify the diffusion and thus act as a Brownian ratchet: as soon as a given length of the biomolecule exits through the pore, a protein binds to it and prevents its diffusion backwards. The dynamics of translocation is typically modeled by a diffusion equation for the length $x$ of the molecule that has already passed through the pore

$$\frac{\partial P}{\partial t} = \frac{\partial}{\partial x} D \left[ \frac{1}{k_B T} \frac{\partial E(n)}{\partial x} P + \frac{\partial P}{\partial x} \right] \quad (46)$$

Here, $D$ is the diffusion coefficient of the biomolecule and $E$ represents the potential (of energetic or entropic origin) through which the biomolecule diffuses. To analyze in detail the process of translocation of nearly stiff biomolecules in the presence of binding proteins, Brownian Dynamic simulations were performed in Ref. [34]. It was found that particles that bind reversibly to the chain give rise to a net force that pulls the chain into the cell significantly faster than pure or even ratcheted diffusion. But it was also found that there are substantial nonequilibrium effects in the dynamics of the translocation. The force that pulls the biomolecule depends strongly on how fast the translocation occurs compared to the binding of proteins. If both processes occur at a similar time scale, the simple Eq. (46), where $E$ represents the potential corresponding to equilibrium adsorption of the proteins, was not able to describe accurately the process. The discrepancies observed justified the need of considering the diffusion process in an extended space in which the dynamics of binding plays an important role.

The nonequilibrium dynamics of translocation of a stiff chain in the presence of binding particles was successfully described using MNET [34]. When translocation is fast compared to the time it takes for the proteins to bind, the dynamics of binding plays a very important role. One may then consider as the $\gamma$ variables in this case both the length of the chain which has passed through the hole $x$, and the number of proteins attached to it, $n$. The dynamics of traslocation is thus considered as a coupled diffusion process in $(x,n)$-space. One can then follow the steps indicated in Section 4 to obtain the Fokker-Planck equation governing the evolution in time of the probability density $P(x,n,t)$

$$\frac{\partial P}{\partial t} = \frac{\partial}{\partial x} D_{rod} \left[ \frac{1}{k_B T} \frac{\partial A(x,n)}{\partial x} P + \frac{\partial P}{\partial x} \right] + \frac{\partial}{\partial n} D_n \left[ \frac{1}{k_B T} \frac{\partial A(x,n)}{\partial n} P + \frac{\partial P}{\partial n} \right] \quad (47)$$

which provides a complete description of the kinetics of both chain entry and particle binding. Here $A$ is the free energy, $D_{rod} = k_B T / \zeta_{rod}$ is the spatial diffusion coefficient of the rod, with $\zeta_{rod}$ the corresponding friction coefficient, and $D_n$ is the kinetic rate constant for the process of particle binding and unbinding, which can be approximated by the expression $D_n = acD_0$, where $a$ is a length of order the particle size, $c$ is the concentration of the binding particles and $D_0$ their spatial diffusion coefficient, obtained from the Smoluchowski theory of aggregation dynamics. The mean first-passage time, the mean force and the average number of proteins attached to the chain can be computed from the Langevin equations related to the Fokker-Planck equation. The results for the average translocation force are represented in Fig. 5 for two different situations corresponding to fast and slow chain entry. They agree with those obtained by means of Brownian molecular dynamics simulations.

When the kinetics of binding is very fast compared to the translocation, one then reaches an instantaneous equilibrium adsorption corresponding to any given length $x$, and the fast variable $n$ can be eliminated from Eq. (47). The resulting equation is then Eq. (46) which holds for equilibrium adsorption. This example reinforces the importance of considering all the nonequilibrated variables for a proper description of the dynamics of an out-of-equilibrium system, and the success of the MNET framework based on restoring local equilibrium in an extended space of variables.
The procedure described previously has also been used to analyze the role played by translational and rotational degrees of freedom of the clusters in the nucleation kinetics [35]. It has been shown that the nucleation rate is greatly influenced by the dynamics of those degrees of freedom and that its expression differs from that obtained when the cluster is considered at rest and only the number of its constituents particles is taken into account. Experimental results [36] have corroborated this more complete scenario.

## *3. Local equilibrium in slow relaxation systems*

In the previous subsection we have seen how the local equilibrium concept depends on the set of thermodynamic variables that are used. Local equilibrium can be recovered by increasing the dimensionality in the variable space where the process takes place. Remarkably, certain features which are considered as new and striking behaviors of nonequilibrium systems, such as the violations of the fluctuation-dissipation theorem and the Stokes-Einstein relation, are the result of a lack of completeness in the description of the processes. They can easily be explained starting from the MNET description at local equilibrium in the extended space and then reducing the number of variables of the system.

Consider a system with different characteristic time and length scales whose energy landscape exhibit many local minima separated by potential barriers. The presence of barriers causes slow relaxation of the system. In the two-state model, the minimal relaxation model, one assumes that the relaxation process consists of two main steps: a fast equilibration process in the well followed by a slow relaxation in which the state of the system jumps from one potential well to the other. The states of the system can be parameterized by the values of a reaction coordinate which varies continuously from the initial to the final state passing through a sequence of nonequilibrium states. It is then plausible to assume that the system evolves via a diffusion process in $\gamma$-space. Assuming local equilibrium in $\gamma$-space, variations of the entropy related to changes in the probability density are given through the Gibbs equation (18). Following the steps indicated in Section 4 we will then arrive at the corresponding Fokker-Planck equation, similar to Eq. (16) and from it to its associated Langevin equation

$$\frac{d\gamma}{dt} = -\phi'(\gamma) + J^r \qquad (48)$$

where $J^r$ is a random contribution to the diffusion current, which has zero mean and satisfies the fluctuation-dissipation theorem

$$<J^r(\gamma,t)J^r(\gamma',t')> = 2D<P(\gamma,t)>\delta(\gamma-\gamma')\delta(t-t'). \qquad (49)$$

The fact that in many instances the time of the intra-well relaxation is much smaller than that of the inter-well relaxation is used to justify the reduction of the number of variables of the system by eliminating the fast variables and by keeping just the values of the populations at each well, namely $n_1$ and $n_2$. One then says that the system evolves via activation. Under this approximation, the dynamics of the system can be described through the kinetic equation [37]

$$\frac{dn_1}{dt} = -\frac{dn_2}{dt} = k_\leftarrow n_2 - k_\rightarrow n_1 - J^r(t) \qquad (50)$$

where $J^r$ is the random current in the new description whose correlation is given by
$$<J^r(t)J^r(t')>=(k_{\rightarrow}<n_1>+k_{\leftarrow}<n_2>)\delta(t-t') \quad (51)$$
which clearly shows that the correlation is not given by a fluctuation-dissipation theorem. Only when fluctuations take place around equilibrium states, in which detail balance holds $(k_{\rightarrow}n_1^{eq}=k_{\leftarrow}n_2^{eq})$, the previous relation reduces to the fluctuation-dissipation theorem
$$<J^r(t)J^r(t')>_{eq}=2k_{\rightarrow}n_1^{eq}\delta(t-t') \quad (52)$$
We can then conclude that the drastic elimination of the fast variables makes the fluctuation-dissipation relation break down. In contrast, it holds when both relaxation processes are considered. Following similar arguments one can show that the Stokes-Einstein relation, in which the diffusion coefficient is proportional to the inverse of the viscosity, does not apply to supercooled colloidal suspensions with slow relaxation [38, 39].

The existence of a fluctuation-dissipation relation, according to the previous result, relies on an equilibrium state [40], or more generally, on a local equilibrium state. In fluctuating hydrodynamics theory, the fluctuation-dissipation theorem is assumed to have the same form as in equilibrium but with the local temperature replacing the equilibrium temperature. Its validity has been corroborated in light scattering experiments in a fluid under a temperature gradient at local equilibrium [41,42, 43]. In any other case, the fluctuation-dissipation theorem is not fulfilled [44]. This holds even in simple systems subjected to an external driving force, as a Brownian particle in a periodic potential [45] or in a shear flow [46]. Recent investigations in systems with memory [47] have established a hierarchical connection between mixing, the ergodic hypothesis and the fluctuation-dissipation theorem [48]. The validity of a fluctuation-dissipation relation, its experimental verification, and its connection with mixing have been reviewed in [49] and [50].

## 4. Nonequilibrium temperatures

The theory presented enables us to analyze the meaning of a temperature in situations far away from equilibrium. To this end let us consider the case of a diffusion process in the presence of inertial effects, as discussed in 6.1. In such a process, the temperature $\tilde{T}(x,v)$ at which the entropy production would be zero is given by [30]
$$\frac{1}{\tilde{T}(x,v)}=-\frac{1}{vm}\left(k_B\frac{\partial \ln P(x,v)}{\partial v}\right) \quad (53)$$
This equation can be rewritten in a form similar to that of the equilibrium temperature:
$$\frac{1}{\tilde{T}(x,v)}=\frac{\partial s_c(x,v)}{\partial e(v)} \quad (54)$$
where $s_c=-k_B \ln P(x,v)$ is the configurational entropy and $e(v)=mv^2/2$, the kinetic energy. The definition of an effective temperature $\tilde{T}(x,v)$ is, however, not unique. If we take $e(v-\bar{v})$ instead of $e(v)$, with $\bar{v}$ the average velocity, the resulting temperature would be that of local equilibrium. This local equilibrium temperature will give a nonzero entropy production. In general, because $\tilde{T}(x,v)$ is a function of both $x$ and $v$, given a point $x$ in space, there is no temperature at which the system would be at equilibrium, i.e., $\tilde{T}(x,v) \neq \tilde{T}(x)$. If an effective temperature at a point $x$ were defined, it would depend on the way the additional coordinate is eliminated. Thus, ambiguities in far-from-equilibrium quantities arise when considering a lower dimensional space than the one in which the process is actually occurring. This is to some extent similar to what happens with effective temperatures defined through fluctuation-dissipation theorems. In such cases, the effective temperature can depend on the scale of observation. It is interesting to point out that all of these effective temperatures, despite their possible analogies with the equilibrium temperature, do not have to follow the usual thermodynamic rules because the system is not actually at equilibrium at temperature $\tilde{T}(x)$. The previous interpretation of an effective temperature is consistent with its characteristics obtained from the two-state model discussed in 6.3. For an activated process in which the description is performed in terms of the initial and final state, the temperature inferred from a fluctuation-dissipation relation is not a robust quantity because it depends on the observable and on the initial conditions of the system [37]. Thus, a possible thermodynamics in which thermal effects are characterized by effective temperatures would not have a consistent formulation.

# 7. Additional applications and comparison with other theories

MNET provides a thermodynamic basis to the stochastic dynamics of systems outside equilibrium and shows that thermodynamic concepts can be applied at mesoscopic scales, where it was believed that thermodynamic arguments were of no use. MNET gives information not only about the evolution of the probability distribution function but also about the thermal and mass exchange processes between the system and the bath. These processes are subject of a great interest nowadays and are being intensively studied in small systems with the purpose of analyzing the validity of the thermodynamic concepts in situations in which the fluctuations become very important. The kinetic equations for systems as clusters, single molecules, or mesostructures can be obtained through MNET in very general situations, including the ones in which the system is subjected to driving forces or gradients or those in which it evolves through potential barriers of any kind.

*1. Additional applications*

MNET has also been applied to other situations involving systems of very different nature. In the next paragraphs we will discuss very briefly the main achievements.

*Activated dynamics*

As shown in 5.2, MNET makes it possible to account for the intrinsic nonlinear behavior of the activated dynamics. In the study of activated process, it has been used to derive the Butler-Volmer equation of electrochemistry [51], originally formulated under empirical grounds. MNET has also been used to study phase transitions at interfaces, such as evaporation and condensation phenomena [52], and to analyze active transport through biological membranes in which ions take large amounts of energy from the hydrolysis of ATP to move from low to high concentrations trough protein channels [53]. In this process, the diffusion current is a nonlinear function of the chemical potential differences of the ions at both sides of the membrane.

*Aggregation and growth phenomena*

MNET has also been applied to the study of growth phenomena driven by surface tension effects [54]. MNET provides expressions for the growth rates which depend on geometric parameters of the aggregates, such as their volume or surface, or on the number of the constituent single particles [55,56]. They are obtained from the corresponding entropy production in the space spanned by those parameters. An interesting example also involving nucleation phenomena at early stages is the case of polymer crystallization [57].

*Nonequilibrium steady states*

MNET has been used to analyze the peculiar characteristics of systems at nonequilibrium steady states [58], providing a mesoscopic thermodynamic framework from which one derives equations of state of systems outside equilibrium [59]. A statistical mechanical model leading to the existence of a stationary state was proposed in [60]. The pressure of a sheared suspension of Brownian particles obtained from the kinetic part of the pressure tensor shows a nonanalytical dependence on the shear rate. Its form is the same as that previously obtained for liquids from projection operators [61] and from kinetic theory [62,63]. The thermodynamics of nonequilibrium steady states of single Brownian macromolecules have been investigated in [64] also by means of an entropy production, leading to the formulation of a mesoscopic theory for single macromolecules consistent with the second law. A steady-state thermodynamics for molecular motor proteins has been formulated in [65]

*Quantum systems*

MNET has also been brought to the quantum regime. MNET provides a master equation for the density matrix of the system, which is obtained from the entropy production in a similar way as in the classical case. It has been applied to spin systems to rederive the Bloch equations and to harmonic oscillators to obtain a quantum mechanical Langevin equation [66]. The formulation of the theory in the quantum domain is useful to analyze the properties of nonequilibrium quantum systems and to elucidate the role played by dissipation in their evolution.

*2. Comparison with other theories*

There are many situations in which the general results obtained with MNET encompass the results derived by other approaches, such as kinetic and stochastic processes theories and projection operator techniques. Our purpose in this section is to discuss some of these situations showing the generality of the MNET formalism.

*Kinetic theory*

MNET provides in general kinetic equations of the Fokker-Planck type. The Fokker-Planck equation obtained with MNET for a dilute suspension of Brownian particles under a temperature gradient [67] coincides with that derived from kinetic theory [68,69]. The case of a velocity gradient has also been studied [46]. Those equations can also be obtained with MNET for higher concentrations when direct and hydrodynamic interactions are relevant [70,71], as in the case of semidilute and concentrated polymer solutions [72]. Hydrodynamic interactions are introduced through the matrix of Onsager coefficients, which is proportional to the Oseen tensor. Direct interactions come into the description through a contribution to the chemical potential. For polymers in the concentrated regime, the Fokker-Planck equation expressed in terms of the monomer concentration field [72] has the same form as that proposed in the classical monograph of Doi and Edwards [73]. For a Brownian particle moving in a granular flow [74] in a homogeneous cooling state, the resulting Fokker-Planck equation is the same as that derived from the kinetic of gases with inelastic collisions [75].

*Stochastic processes*

The scheme presented provides a phenomenological procedure to derive the Fokker-Planck equation describing the dynamics of mesoscopic systems [10]. This formalism can be useful in cases in which the complexity of the system or the nonequilibrium nature of the environment makes a detailed description of the problem impractical. Representative examples are the kinetic processes discussed in 5.3 and 6.2. The theory also applies to nonlinear transport systems in the presence of memory effects which are introduced through the time dependence of the Onsager coefficients [76]. In these cases the Fokker-Planck equations have the same form as those obtained from Langevin and master equations. As an example, the resulting generalized Fokker-Planck equation, related to the generalized Langevin equation, coincides with that obtained in [77] for non-Markovian systems with Gaussian noise.

*Projection operators*

Brownian motion in the presence of external gradients has also been studied by means of projector operators [78]. For the case of a temperature gradient, the Fokker-Planck equations obtained with projector operators [78] and MNET [67] are identical. Nonlinear Langevin equations for the moments of the distribution function can be derived from the Fokker-Planck equation obtained with MNET when the Onsager coefficients depend on the state variables [76]. The resulting equations coincide with those obtained from projector operators [79].

*Langevin descriptions*

The relationship between the Langevin equation and the laws of thermodynamics was discussed in [80]. This formalism, known as stochastic energetics, has been applied to energy transduction processes and to the characterization of nonequilibrium steady states [81]. A mesoscopic nonequilibrium thermodynamic formalism for the Langevin equation has also been proposed and applied to single macromolecules [64,82]. In the approach we have presented, the Langevin equation is straightforwardly connected to thermodynamics through its Fokker-Planck equation.

## 3. Advantages of MNET

In the previous paragraphs, we have shown that MNET encompasses the results obtained by means of other more complex nonequilibrium statistical mechanics theories. A distinctive feature of MNET is that it provides a straightforward formalism to implement the dynamics of non-equilibrium systems from the knowledge of their equilibrium properties. The advantages of using MNET become especially manifest when there are dynamic processes of different nature taking place simultaneously, as for instance, when the system is subjected to fluctuations and exchanges heat or mass with a nonequilibrium environment that has its own dynamics. In such cases, the Langevin and the Fokker-Planck equations are not a mere extension of those formulated for simpler situations and must be derived by means of a nonequilibrium statistical mechanics theory. MNET uses a systematic and simple method through which those equations can easily be obtained. The cases of the translocation of a biomolecule and of the Brownian motion in a nonequilibrium fluid discussed previously are illustrative examples. In the former, the Fokker-Planck equation contains two currents corresponding to the two relevant dynamic variables. In the latter, the imposed gradient not only affects the intensity of the noise but also the form of the kinetic equation by adding a new term that is responsible for thermal diffusion.

To illustrate explicitly the influence of a nonequilibrium environment in the dynamics of the system, we consider Brownian motion in a temperature gradient. The effects of the gradient on the probability current of a Brownian particle can directly be inferred from the entropy production in the space of mesoscopic variables by taking into account its Onsager coupling to the heat current. The form of the probability current is

$$\vec{J}_u = -L_{uT}\nabla T/T^2 - k_B L_{uu}\frac{\partial}{\partial \vec{u}}\ln(P/P_{l.eq}) \qquad (55)$$

where $\vec{u}$ is the velocity of the Brownian particle, the $L$ terms are Onsager coefficients and $P_{l.eq.}$ is the local equilibrium distribution function [67]. The presence of particles, in turn, modifies the heat current through the system. This effect can also be analyzed through the entropy production. The resulting heat current is

$$\vec{J}_q = -L_{TT}\nabla T/T^2 - k_B \int L_{Tu}\frac{\partial}{\partial \vec{u}}\ln(P/P_{l.eq})d\vec{u}, \qquad (56)$$

where the Onsager coefficients obey the Onsager relation $L_{Tu} = -L_{uT}$. These equations clearly show the existence of a coupling between the two irreversible processes present in the system: probability diffusion and heat conduction. The resulting Fokker-Planck equation

$$\frac{\partial P}{\partial t} = -\vec{u}\cdot\nabla P + \beta\frac{\partial}{\partial \vec{u}}\cdot\left(P\vec{u} + k_B T\frac{\partial P}{\partial \vec{u}}\right) + \frac{\gamma}{T}\frac{\partial}{\partial \vec{u}}\cdot P\nabla T, \qquad (57)$$

with $\beta$ being the friction coefficient of the particles and $\gamma$, a coefficient related to the Onsager coefficient $L_{uT}$, coincides with that obtained from kinetic theory [68]. The Fokker-Planck equation and the evolution equation for the temperature field provide a complete description of the heat exchange process in the system. This example illustrates the way in which MNET can systematically be used to analyze heat exchange processes between the system and its environment in the presence of fluctuations.

# 8. The actual meaning of being "far away from equilibrium"

The presence of unbalanced thermodynamic forces moves the system away from equilibrium. How far can these forces move the system away from equilibrium, as discussed in the classic monograph [2], depends not only on the values of the force but also on the nature of the process. For transport processes in simple systems, such as heat conduction (Fourier law) and mass diffusion (Fick law), local equilibrium typically holds even when the systems are subjected to large or even very large gradients [3]. We have seen in section 4 that the probability current also obeys a linear law: the Fokker-Planck equation is linear and describes situations that can be far from equilibrium. Linearity does not necessarily imply in those cases closeness to equilibrium. On the contrary, for the wide class of activated processes discussed in Section 6, linearity breaks down already at small values of the affinity, which seems to imply that local equilibrium is lost almost immediately.

The results of the method we have presented indicate that existence of local equilibrium depends on the set of variables used in the characterization. As shown in 5.2, when an activated process is described not just in terms of the initial and final states but through its reaction coordinate, local equilibrium holds. Increasing the dimensionality of the space of thermodynamic variables, by including as many dimensions as nonequilibrated degrees of freedom, leads to local equilibrium in the enlarged space and allows the use of nonequilibrium thermodynamics at shorter time scales in which fluctuations are still present. We can thus conclude that many kinetic processes, such as nucleation, chemical reactions or active transport, which have been assumed to be far away from equilibrium because of their intrinsic nonlinear nature, take place at local equilibrium when a finer description is adopted.

*1. An example: single macromolecule*

To further illustrate how systems brought outside equilibrium may be considered at local equilibrium in an extended space, we will study the case of a macromolecule in a solvent at constant temperature subjected to an external driving force. In addition to the position of its center of mass $x$, the macromolecule is characterized by an additional fluctuating variable $\theta$, which might represent, for instance, its size or its orientation. For small values of the force, local equilibrium in $x$-space holds in such a way that we can formulate the Gibbs equation expressed now in differential form

$$Tds(x) = -\mu(x)d\rho(x) - Fd\Theta(x) \qquad (58)$$

where $F$ is the force and $\Theta(x)$ is the average value of the $\theta$ variable defined as

$$\Theta(x) = \int \theta P(x,\theta) d\theta \qquad (59)$$

with $P(x,\theta)$ being the probability distribution and $\mu(x,\theta)$, its conjugated chemical potential. Let us now assume that the driving force increases in such a way that the system is no longer in local equilibrium in $x$-space. The way to restore local equilibrium is to increase the dimensionality by considering the fluctuating variable $\theta$ as an independent variable and defining the Gibbs equation as

$$Tds(x,\theta) = -\mu(x,\theta)dP(x,\theta). \qquad (60)$$

Proceeding as in section 6, one could obtain from this equation the corresponding Fokker-Planck equation, which would describe the dynamics of the macromolecule in the extended space.

## 9. Conclusions

A typical way to study nonequilibrium mesoscopic systems is to use microscopic theories and proceed with a coarse-graining procedure to eliminate the degrees of freedom that are not relevant to the mesoscopic scale. Such microscopic theories are fundamental to understand how the macroscopic and mesoscopic behavior arise from the microscopic dynamics. On the downside, they usually involve specialized mathematical methods that prevent them to be generally applicable to complex systems; and more importantly, they use much detailed information that is lost during the coarse-graining procedure and that is actually not needed to understand the general properties of the mesoscopic dynamics.

The approach we have presented here starts from the mesoscopic equilibrium behavior and adds all the dynamic details compatible with the second principle of thermodynamics and with the conservation laws and symmetries that are present in the system. Thus, given the equilibrium statistical thermodynamics of a system, it is straightforward to obtain Fokker-Planck equations for its dynamics. The dynamics is characterized by a few phenomenological coefficients, which can be obtained for the particular situation of interest from experiments or from microscopic theories, and describes not only the deterministic properties but also their fluctuations.

We have shown explicitly the applicability of these mesoscopic thermodynamics methods to a broad variety of situations, such as activated processes in the nonlinear regime, inertial effects in diffusion, and transport in the presence of entropic forces. It is important to point out that there are many mesoscopic systems that have been studied with the methods of statistical thermodynamics [5]. The dynamics of most of those systems still remains poorly understood. Our approach opens the way to study their dynamics in terms of kinetic equations of the Fokker-Planck type.

## Acknowledgments

We would like to thank P. Mazur, H. Reiss, D. Bedeaux, S. Kjelstrup, J. Dufty, M. Lozada, A. Gadomski, F. Oliveira, A. Pérez-Madrid, I. Pagonabarraga, G. Gomila, and I. Santamaría-Holek, for fruitful discussions over the years. This work has been partially supported by the DGiCYT of the Spanish Government under Grant No BFM2002-01267.

# List of Figures

**Figure 1:** a) Pore geometries for OmpF porin (the grayish structure embedded in lipid bilayer) and $\alpha$-toxin channel (dots extending to the membrane- bathing solution) from Ref. [14]. b). Steady state current of particles through a 3D hyperboloidal cone (inset) as a model channel (see Ref. [12]). The parameter $\eta_0$ characterize the aperture of the hyperboloidal cone. The dashed line is the exact solution of the 3D diffusion equation, and the heavy line is the result of MNET (Eq. (25)).

**Figure 2**: Potential barrier as a function of the reaction coordinate. $\gamma_0$ indicates the location of the top of the barrier (transition state).

**Figure 3**: Polarizing optical micrographs of poly(ethylene terephthalate) (PET) crystallized at 240°C in the absence (A) and in the presence (B) of a shearing. As a consequence of the shearing nucleation becomes increasingly profuse, and the shape of spherulites becomes elliptical. (From Ref. [22]).

**Figure 4**: Probability distribution function (solution of Eq. (43)) when a concentration gradient is applied (see Ref. [25]). At small values of the velocity relaxation time $\tau$, the distribution function is a Gaussian (full circles); when the relaxation time increases, the distributions shows a non-Gaussian behavior (hollow circles). The vertical dotted line and arrows are guides to the eye to emphasize the asymmetric form of the velocity profile for $\tau = 10$. All values are given in arbitrary units.

**Figure 5:** (a) Snapshot of the Brownian dynamics simulation of the translocation of a rigid chain into a spherical cell. The dark spheres depict the monomers of the chain, the grey spheres are the free proteins, whereas the light grey spheres are the bound proteins. (b) Force driving the translocation as a function of the length of the chain $x$ inside ($\sigma$ is the size of the monomers of the chain). The squares and the hollow circles are the results of simulations for N=100 binding proteins and two different values of the diffusion coefficient of the rod. The heavy and the dashed lines are the predictions of MNET for these values of $D_{rod}$. Note that for small $D_{rod}$ (i.e. slow entry of the chain, which facilitates the equilibration of binding), the translocation force is roughly 4 times larger than for the other case, where translocation and binding occur at the same time scale.

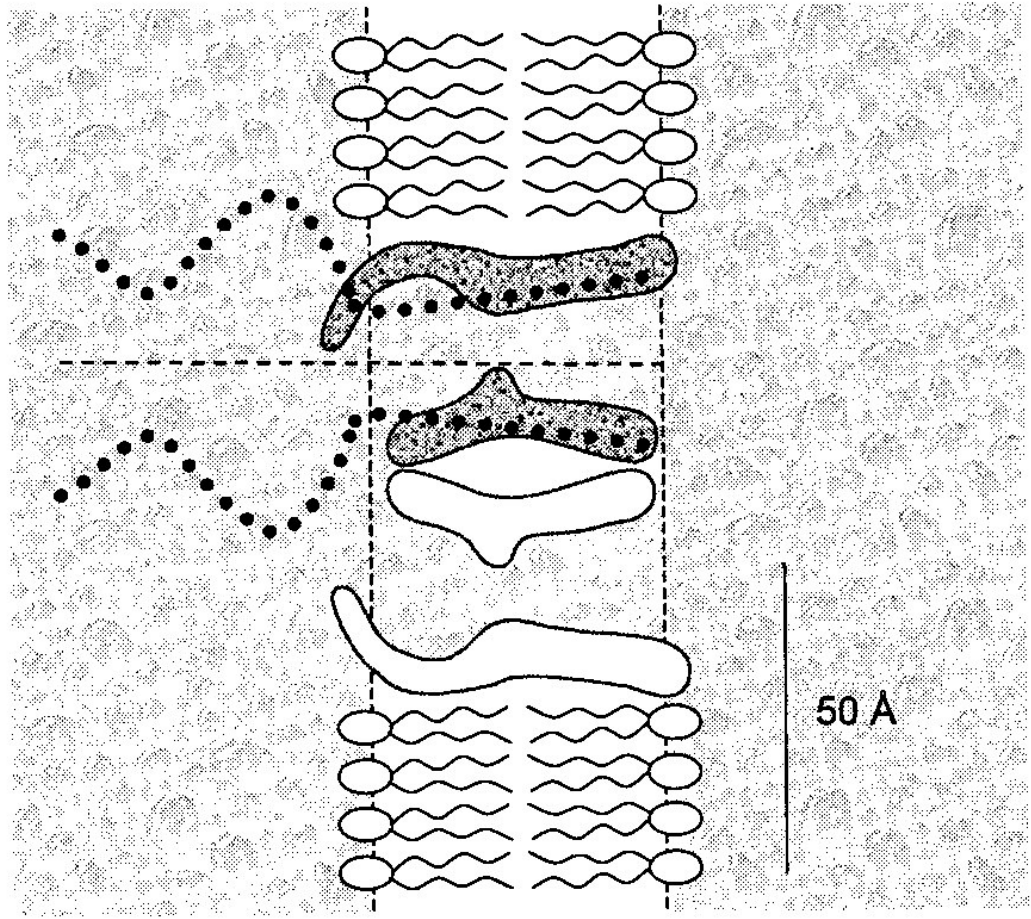

Figure 1 (a)

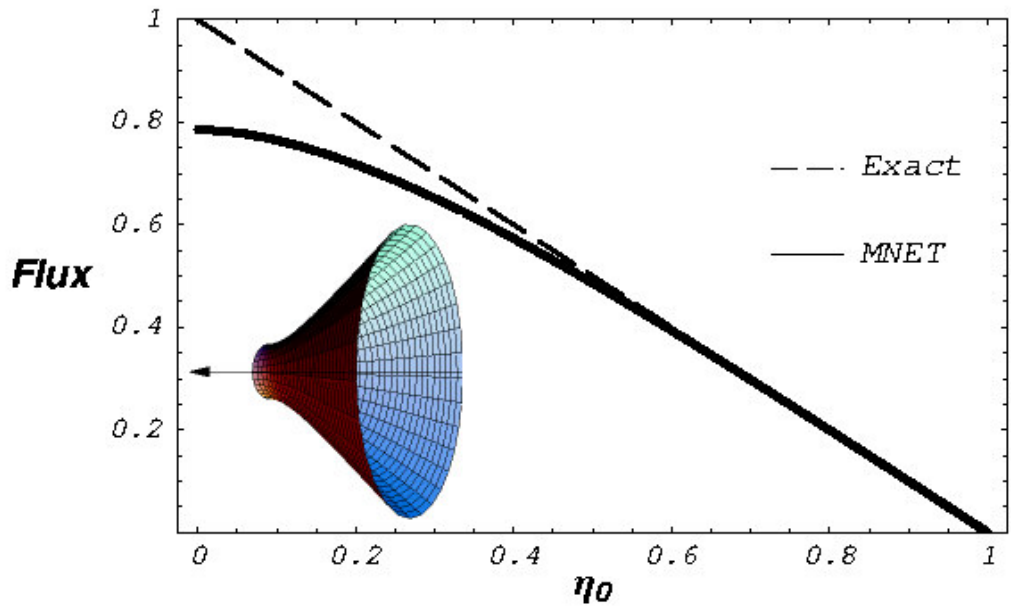

Figure 1b.

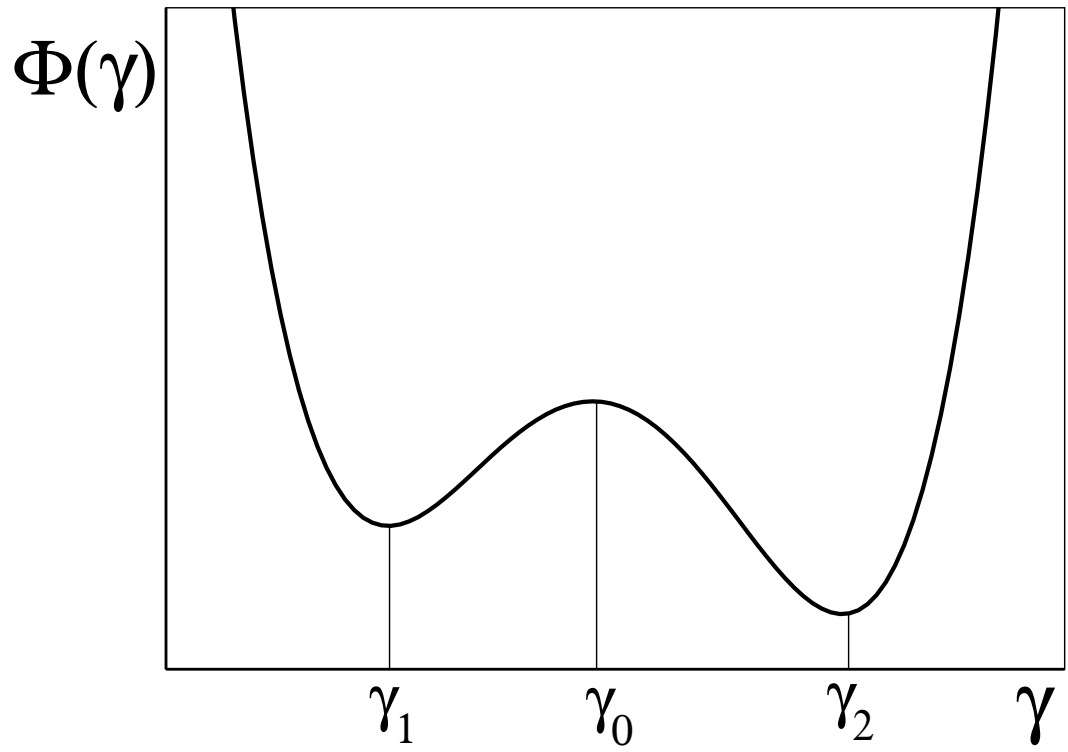

Figure 2.

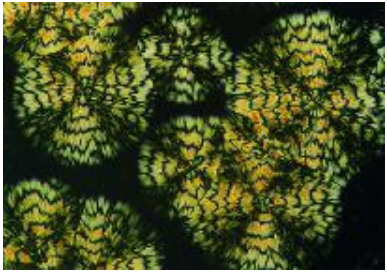

Figure 3(a)

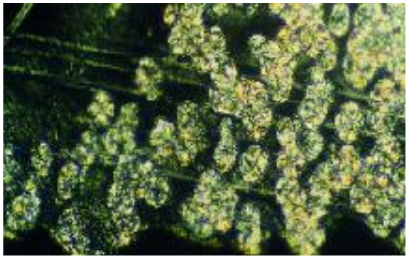

Figure 3(b)

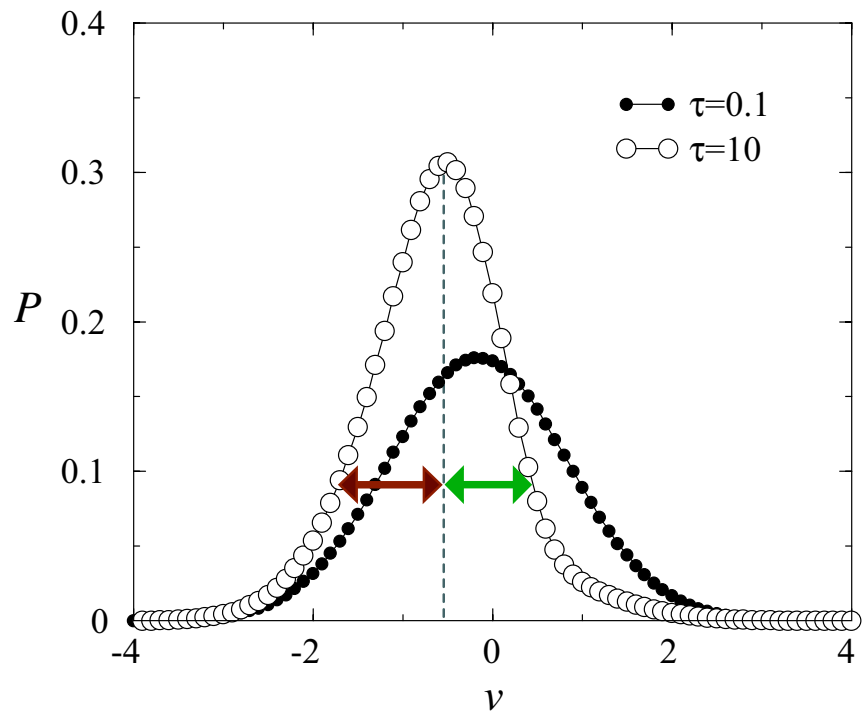

Figure 4.

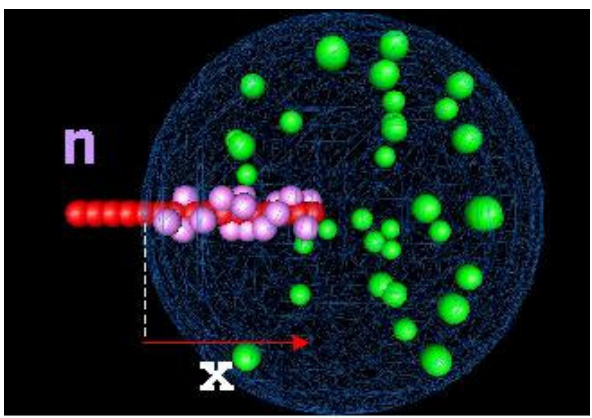

Figure 5(a)

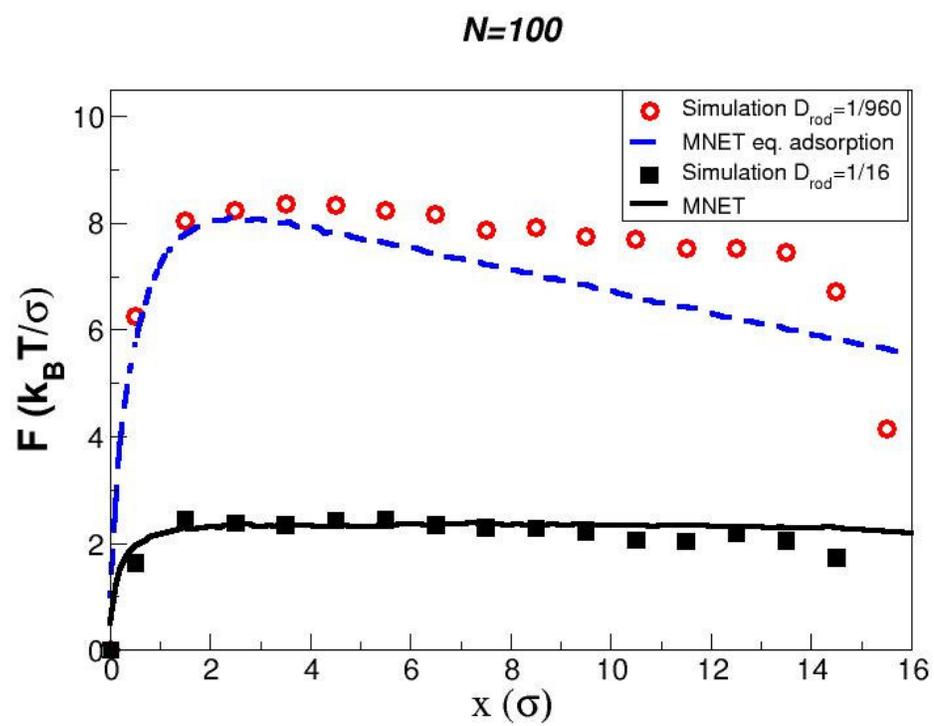

Figure 5(b)